\documentclass[]{aa}

\def\hang{\hangindent\parindent}
\def\textindent#1{\indent\llap{#1\enspace}\ignorespaces}
\def\litem{\par\hang\textindent}

\def\be{\begin{equation}}
\def\ee{\end{equation}}
\def\bea{\begin{eqnarray}}
\def\eea{\end{eqnarray}}

\renewcommand{\thesection}{\arabic{section}}
\renewcommand{\thesubsection}{\arabic{section} \arabic{subsection}}

\usepackage{graphicx}
\usepackage{amssymb}

\begin{document}

\title{Gamma-ray emission from galaxy cluster outskirts versus radio relics}

\author{G. Siemieniec--Ozi\c{e}b{\l}o
\and Z.  A.~Golda
}

 \institute {Astronomical Observatory, Jagiellonian University,  ul. Orla 171, 30--244 Krak\'ow, Poland\\
\email{e-mail: grazyna@oa.uj.edu.pl; zdzislaw.golda@uj.edu.pl}
}

\abstract{
Galaxy cluster peripheries provide important information on the nature of ICM/IGM linkage. In this paper we consider potential future observations in the gamma-ray domain at cluster edges involving the radio relic phenomenon.
}
{
We focus on the spectral signature of gamma radiation that should be evident in the energy range of Fermi--LAT, i.e. $\gtrsim 10^{-1}$~GeV and the CTA energy range $\sim$~$ 10^{2}$~GeV. The spectral signature results from a comparable gamma-ray flux due to the IC and $ \pi ^{0} $ decay on the edge of the cluster, and its spectral position is a function of the magnetic field and relative efficiency of the acceleration of protons and electrons. We aim to draw attention to the dependence of the gamma-ray structure on the magnetic field value.
}
{
As an example, we carried out analyses of two types of non-thermal diffuse radio emission: the radio relic of A~2256 and the radio halo of Coma cluster. We suggest that in both cases the expected spatially correlated gamma-ray spectrum should have a characteristic structure that depends on the strength of the local magnetic field. In both of the clusters we calculated the combined flux of gamma radiation from the actual observational values of the used observables.
}
{
The revealed spectral dependence on the magnetic field would allow us to relate the future spectral observations, in particular the position of the gamma-ray signature, to the value of the magnetic field in the border area between galaxy clusters and their connecting filaments,
 possibly constraining the estimated relative efficiency of particle acceleration at the edge of the cluster.
}{}


\maketitle

\section{Introduction}

Clusters of galaxies are excellent laboratories for studying plasma physical processes. Combining cluster medium observables from the radio to gamma-ray domains enables us to probe the characteristics of intracluster plasma. The intracluster medium (ICM)  is of interest for a number of reasons. With respect to the cosmological aspect, the most important things to understand concern the cluster formation and their subsequent evolution.\\
 First, the important point here is  the role of centrally located black holes and the processes underlying specific heating and cooling of nuclear regions of clusters. 
Second, the baryonic matter deficit in galaxy clusters is another issue to be debated. In practice, it reduces to the  search for  baryonic matter vestiges at the cluster outskirts in the form of cooler gas of the warm-hot intergalactic medium (WHIM). 
Third, the dynamic state of baryons in galaxy clusters and connecting filaments is in turn associated with the shock waves and accompanying turbulences occurring there. The role of these phenomena in the modelling of intergalactic cluster media is one of the motivations for studying shocks.

The main astrophysical motivation is to explain the whole variety of non-thermal and the thermal processes  occurring at different spatial scales of clusters in both
  central regions, where the interaction of AGN with the cluster medium is of primary importance,  and other places, particularly at the periphery, with conditions favouring multi-frequency emissions.

We expect that the above interrelated approaches will finally yield a coherent picture of the ICM. This will allow us to answer  the following key questions  concerning the baryon fraction of clusters, which could be used as an estimator of the average baryon value for the universe and to address other primordial problems: What causes the non-thermal high-energy cluster emission in the radio and hard X-ray domains? What is the origin of the large-scale magnetic field and its long-standing evolution, and, particularly, its coupling to cosmic-ray physics, i.e.  particle diffusion and transport (viscosity, conduction)? What is our    understanding  of the role and properties of turbulences within the ICM?

The presence of relativistic particles in the ICM manifests itself as diffuse radio emission in the form of cluster radio halos and relics. Natural tools for the study of these phenomena are synchrotron diagnostics as well as X-ray and, perhaps soon, gamma analyses.

The peripheral cluster regions in the context of the above questions have a particular significance. On the one hand, investigating cluster outskirts is very important because we can see a connection of clusters with the cosmic web via accretion of gas and subclusters, thus providing a viable cosmological probe. On the other hand, in these outer regions,  closer to the sources of accretion from filaments, it is in a sense easier to study the physical properties of the ICM (e.g.~non-thermal interactions), since they are much less complex than the central regions. 

Recently, several informative observations have been made at cluster peripheries. The baryon content in outer regions has been estimated through Sunyaev--Zel'dovich effect and X-ray observations (e.g.~Afshordi et al.~\cite{Afshordi}). For a number of galaxy clusters, measurements of intracluster gas temperature and entropy profiles, along with profiles of gas density, gas fraction, and mass, out to large radii, have also been performed (e.g.~Reiprich et al.~\cite{Reiprich}; George et al.~\cite{George}; Sato et al.~\cite{Sato}) based on SUZAKU data.

Signatures of giant shock waves seen as radio relics and arising from cluster mergers were discovered in the last years (e.g.~Bagchi et al.~\cite{Bagchi}; van Weeren et al.~\cite{Weerena}; van Weeren et al.~\cite{Weerenb}). The recent  discovery of large-scale diffuse, non-thermal radio emission in PLCK G287.0+32.9 (Bagchi et al.~\cite{Bagchi:a}),  which was confirmed by the Planck satellite in  an all-sky blind search for new clusters through the Sunyaev--Zel'dovich effect, reveals a pair of giant ($>1$~Mpc) arcs-shaped peripheral radio relics.

These relic sources constitute unique signs of energetic mergers and shocks and a probe of the filamentary cosmic-web structure beyond the cluster virial radius.
Although currently there is not too much other observational evidence about physical properties within cluster outskirts, numerical simulations tend to describe at least the thermodynamic state of the gas. They predict numerous shocks in the large-scale structure and are eventually able to reproduce the radio luminosity  features through analysing the parameters of the merger shocks during their evolution.

In this paper, we focus on the external cluster features, i.e. locations of radio relics. Below we present the reasons behind our supposition that these radio-emitting, extended boundary areas should also show a certain brightness in gamma emission  within the same timescale as radio emission. Then we are going to show that the spectral characteristics of emission depends on the strengths of peripheral magnetic fields in clusters.

In the frame of the model of radio relics adopted  in the literature, we present in Sect.~2 the arguments in favour of the possibility of the occurrence of external gamma emission that is spatially correlated with the relic; in Sec.~3, we briefly discuss the mechanisms underlying production of gamma rays. In Sec.~4, we present the expected  gamma-ray spectra for two well-known and analysed cases of galaxy clusters.

\section{What we can see and what we expect to see at the edges of clusters}

The extended radio emission regions that can be observed at the edges of many clusters (Hoeft \& Br\"uggen~\cite{HoeftBruggen}; Skillman et al.~\cite{Skillman}; van Weeren et al.~\cite{Weeren}; Bonafede et al.~\cite{BonafedeBruggen}), commonly referred to as relics, are associated with the recently shock-accelerated electrons. The origin of these boundary relativistic electrons is accounted for by the DSA acceleration mechanism in a moving forwards merger shock, which in many clusters can give spectral predictions that are consistent with the observed distribution of radio spectral index across the relic width (e.g.~A~3667).

Below we consider a hypothesis that within the relic extent we can also expect an enhanced gamma emission. Therefore, first we present a possible origin of these gamma-ray, ring-like structures at the edges of clusters.

\subsection{How do the main observables characterizing the relic area behave?}

Very little is known to date about magnetic field properties in the cluster outskirts. However the mean magnetic field spatial profiles presented in the literature, which are beta-modelled  and volume-weighted via synchrotron emission,   (see e.g.~Arieli et al.~\cite{Arieli}) show a decrease of magnetic field strength within the virial radius of more than 2 orders of magnitude. This radial magnetic field distribution roughly agrees with the spatial distribution of the field inferred from RM measurements of a statistical sample of galaxy clusters (Clarke et al.~\cite{Clarke}).
        
Also (Donnert et al.~\cite{Donnert:b},~\cite{Donnert:a}) successfully fitted the predicted rotation measure signal of galaxy clusters to the observations. Their magnetic field profiles are in good agreement with the recent findings, showing that the radial magnetic field decline conforms to the density profile.
        
 Energy density spatial profiles of the gas and cosmic rays (CR) in the ICM were recently discussed in the framework of large-scale cosmological shock simulations (e.g.~Vazza et al.~\cite{Vazza}; Burns et al.~\cite{Burns}); these profiles were also compared with the new observations of Suzaku, out to $r_{\rm vir} \approx r_{200}$ (Walker et al.~\cite{Walker}). The value{ 
$r_{200}$ is the virial radius of a spherically collapsed cluster with a virial density $\approx 200$ times the critical density of the Universe.
}  
 The inconsistency of these profiles with the hydrostatic equilibrium assumption in the outer cluster regions suggests that the ICM pressure does not come solely from the thermal energy density at $r \approx r_{\rm vir}$.

Also the underestimates (up to 30$\%$) of galaxy cluster  masses, which increase with $r$, can indicate the presence of greater amount of gas in the outskirts than it is predicted by an ordinary $\beta$ model. This can result either from  gas or  subcluster accretion along  the  filaments entering clusters in these regions, such as from the existence of  connections with the cosmic web, but also, in the case of an advanced outgoing merger, from the compressed  post-shock gas.  For both cases, the  hydrostatic equilibrium assumption becomes invalid at the outskirts of clusters (Sato et al.~\cite{Sato}). 

Several papers dealing with galaxy cluster merger simulations reveal the final radial profiles of various quantities, as compared to the initial radial profiles. This includes not only the plasma density and temperature profiles but also the velocity and entropy maps. In the latter case, the final profiles exhibit a large increase in entropy in the cluster outskirts, which can indicate the dissipative processes at work there, presumably associated with the strengthening of the shock.

The same trend is confirmed in radio profiles of gas density and in the temperature and entropy profiles recently presented in the paper of Vazza et al.~\cite{Vazza},
 including  CR physics (CR injection,  feedback, and acceleration efficiency).

 The effects of CR input are particularly evident in the advanced phase of merging systems, where pressure of freshly injected CR's at peripheral position can be comparable with the thermal pressure.  The role of CR physics can be crucial in interpreting the recent Suzaku observation of a galaxy cluster close to $r_{200}$ and beyond (Bautz et al.~\cite{Bautz}; Simionescu et al.~\cite{Simionescu}) and in predicting their future non-thermal observations at larger radii.

The statements above and, in particular, the  recent results concerning the dynamical role of CR in cluster outer regions makes us expect a diffusive gamma-ray emission in the same region, i.e.~close to $r_{200}$.
        
On the basis of all the mentioned observational indicators, we assume the following:
        \begin{enumerate}
\item The relic radio features are mainly generated through interior forwards-propagating shocks associated with cluster mergers. This hypothesis is supported by simulations (e.g.~Skillman et al.~\cite{Skillman}), which show that almost every cluster displays signs of radio relics at some stage of its evolution, being highly dependent on the merger, the radio emission of which is associated with the merger shock parameters. However, the radio emission decreases when the ICM  overdensity value drops below $\approx 10 \div 30$, which disfavours the possibility of attributing the relic feature to the accretion shock alone. Therefore we postulate here that the region located  exactly between the outwards-going shock ($\lesssim r_{\rm vir}$) and the accretion shock $r_{\rm acc}$ should be expected to show the multi-frequency, non-thermal emission, such as radio, gamma, and hard X-ray  (HXR).
        
This also means that the relic feature is related to and dependent both on the merger and accretion shocks' properties.
        \item An approaching merger shock together with a long-living accretion shock form a so-called double shock structure (Fig.~\ref{Figure01}), which provides for larger efficiencies both in injection and CR acceleration.
        \begin{figure}[htb]
        \begin{center}
\includegraphics[totalheight=4 cm]{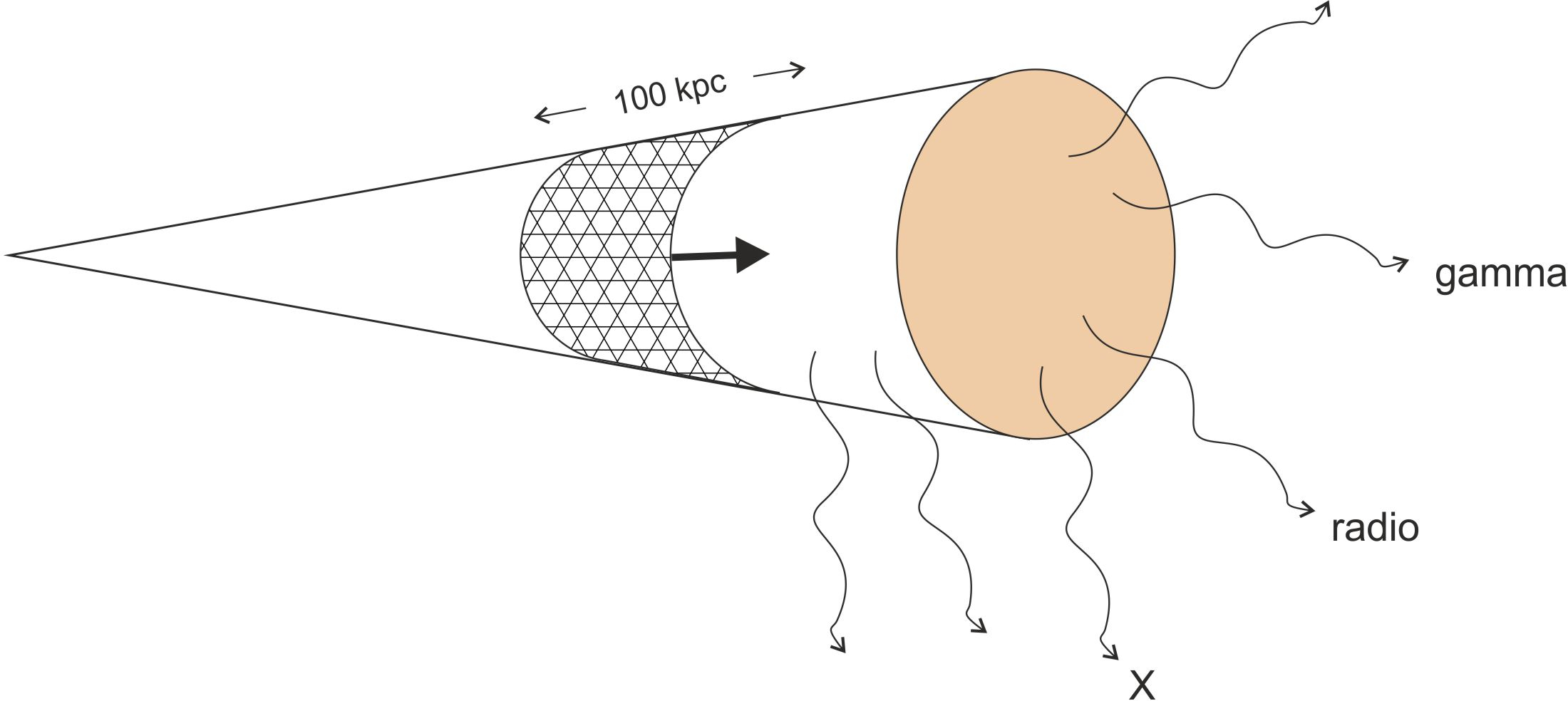}
                \end{center}
\vspace{-4mm}
\caption{Outline of the considered region of radio relic in between the outgoing merger shock and accretion shock  for galaxy clusters.}
\label{Figure01}
        \end{figure}
When pre-shock gas density is slightly higher, which is the case in a  forwarding shock, then the density of accelerated CRs is enhanced. The more considerable gas density at larger radii is due to adiabatic compression interior to accretion shock. This  does not contradict that the ratio $v_{\rm sh}/v_{\rm sound}$  still increases radially when approaching the outer region, in spite of gas redistribution due to the passage of the shock, which is obviously equivalent to a Mach number  
{ 
  ($M = v_{\rm sh}/v_{\rm sound}$)
} 
increase of the merger shock. Apart from the increasing power of forwarding shock, in the vicinity of accretion shock we expect the presence of the older population, such as preexisting CRs, which also increases the acceleration efficiency for moderate $M$ (Hong et al.~\cite{Hong}; Kang et al.~\cite{Kang}).  Thus between the merger and the exterior strong accretion shock, the effective, i.e. combined, acceleration efficiency, must be enhanced and so is the flux of CRs accumulated between both shocks. Such a double shock structure has been previously discussed (Siemieniec--Ozieblo \& Ostrowski~\cite{Siemieniec--Ozieblo}) in the context of UHECR acceleration in large-scale structures, showing a significant amplification of acceleration efficiency.

The main consequence of accelerated particle confinement in two such converging flows is a hardening of the spectral index owing to inefficient particle escape  from the acceleration region and, thus,  to the diffusion enhancement between the shocks (see details in Siemieniec--Ozieblo \& Ostrowski~\cite{Siemieniec--Ozieblo}).

The exact formalism describing development of this double shock model in the current context will be presented in the next paper, which will include the analysis of diffusion of particles between two shocks: a forwarding merger shock and a stationary accretion shock.
        \end{enumerate}

\subsection{Do gamma-ray relics exist at the cluster outskirts?}

Although no gamma-ray emission from the clusters observed by Fermi--LAT (e.g.~Ackermann et al.~\cite{Ackermann}) has been detected to date, it is still expected to occur within the observationally established constraints. The (spatial profiles of) expected morphology and the spectrum of  gamma-ray emission should depend on the relative contribution of two gamma-ray mechanisms dominant at the cluster outskirts. In the photon energy range, where the  gamma-ray emission would have the maximal flux, i.e. $10^{-2} \div 10^{2}$~GeV, both relativistic protons and primary electrons (via IC) should equally reveal their presence (see e.g.~Fig.~1 in Pinzke \&~Pfrommer~\cite{Pinzke}).

Primary CR electrons contribute to an irregularly elongated relic feature seen in radio, HXR, and possibly in  gamma-ray domains, due to synchrotron and inverse Compton (IC) processes. The timescales of these processes are relatively short for the electrons and typical magnetic field of the relic: $t_{\rm radio:synch+IC}\sim 10^{8}$~y, $t_{\rm HXR:IC} \sim 10^{7\div 8}$~y; while   $t_{\rm gamma:IC} \sim 10^{7}$~y.

The long-living  gamma-ray emission that results from CR protons interacting with thermal gas protons is expected to be a dominant component everywhere but in the peripheral regions of the merging cluster, where it can have a value that is comparable to the primary IC emission. The maximum energy attainable by protons ($E \sim 10^{19}$~eV) is determined by the time of the propagating merger shock, which is $\tau_{\rm shock} \approx 10^{9}~{\rm y}$.

All the above timescales can suggest that the relic sources can be treated as a sort of transient ``flares" accompanied by X-ray, gamma, and radio emissions (Fig.~\ref{Figure01}). 
 {  
This simply means that the higher frequency used to observe the IC emission within the relic, the shorter the high-energy phenomenon to be found there corresponding to the shorter cooling time.
} 

As one can see from Fig.~\ref{Figure01}, the possibility of observing a hypothetical gamma emission strongly depends on the direction of merger propagation. The most favourable case is the merger shock propagating close to the line of sight. It is obvious that its face-on geometry increases the detectability of relic radiation.

\section{Gamma-ray spectral properties}

In this paper we consider two instances of such sources: the cluster A~2256 and the Coma cluster as a reference case; the relic associated with the former is caused by a shock moving approximately along the line of sight. In this example, the modelled spectral index for the relic is very flat, suggesting that we observe the advanced phase of merger, in which the radio emission comes from freshly  and efficiently accelerated electrons within the narrow volume around the shock.

In the case of A~2256, we have already proposed a coherent model consistently explaining the radio and HXR emissions (Siemieniec--Ozieblo \& Pasternak~\cite{Siemieniec--OziebloPasternak}). One can find there some predictions concerning the gamma emission owing to $\pi^{0}$-decay. The resulting gamma-ray flux estimated for the A~2256 relic is by $2\div 3$ orders of magnitude smaller than in the case of the halo of the Coma cluster. Now, we additionally calculate the primary IC component since at the cluster outskirts it should equally  contribute to the gamma emission. Moreover, we expect that in the energy range covered by  FERMI, the gamma-ray spectrum can show a specific concave structure in the relic location. This feature would result from comparable IC and $\pi^{0}$-decay emissions in the energy range close to the $\pi^{0}$ bump. 

\begin{figure}[htb]
                \begin{center}
\includegraphics[totalheight=5.5 cm]{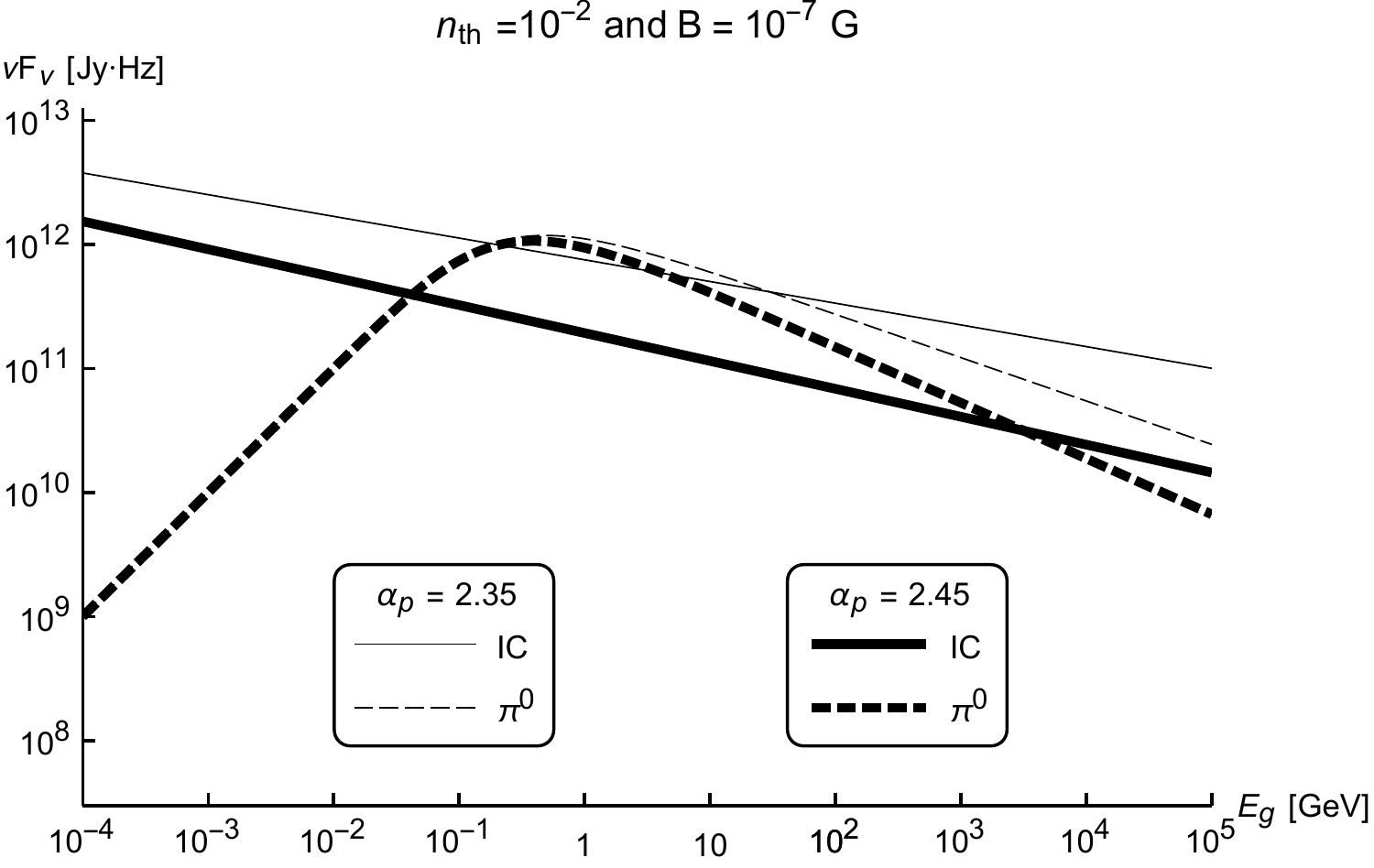}
                \end{center}
\vspace{-4mm}
\caption{Comparison of $E_ {\rm cross}$ dispersion for two slightly different spectral 
 indices $\alpha_{\rm p}$, for the IC  and $\pi^0$ fluxes in a wider energy range.}
\label{Figure02}
        \end{figure}

Potentially, the same characteristic could be produced on both sides of the $\pi^{0}$ bump, i.e. also for $E \geq 10$~GeV. 
Although we focus here on the energy regime accessible to the Fermi--LAT observations, one has to mention that an equally interesting domain with respect to both IC and pion-generated photons is expected at the CTA energy range, i.e. above a few tens of~GeV. 
 However, the observational verification of the expected structure at such a high-energy range significantly depends on the IC spectral index and thus it is distinctly visible only for flat spectra; cf. Fig.~\ref{Figure02}. In the energy range $E \geq 10$~GeV, the location of the feature is extremely sensitive to the energy spectrum. As one can see in Fig.~\ref{Figure02}, the relative change of spectral index $\frac{\Delta \alpha_p}{\alpha_p}=0.04$ corresponds to $\Delta E_{\rm cross}$ of the order of $3 \times 10^{3}$ GeV in the CTA range.  
{  
Another crucial assumption concerns the maximum energy of IC photons reaching the TeV range (e.g.~Miniati~\cite{Miniati}; Kushnir \& Waxman~\cite{KushnirWaxman}), which is connected with the uncertainty of maximum energy to which CR electrons can be accelerated in the double shock considered. We estimate the maximum energy of primary electrons in a standard way, i.e. by comparing the acceleration time, $\tau_{\rm acc}$ ($\tau_{\rm acc}\propto \gamma_{\rm e}$ B$^{-1}$) with the cooling time due to IC and synchrotron processes $t_{\rm cool}$. This allows us to derive the maximal value of Lorentz factor for electrons $\gamma_{\rm e}^{\rm max}$, which constrains the energy break in electron spectrum $\gamma_{\rm e}^{\rm br}\sim\gamma_{\rm e}^{\rm max}$. The outskirt cluster region covered by the merger shock approaching the accretion shock, in which the acceleration occurs, is characterized by strong turbulence. Accordingly, we expect the effective Bohm diffusion assumption to be valid  there. Ultimately, this results in a value of $\gamma_{\rm e}^{\rm br}\sim 10^8$ for a magnetic field B $\sim 0.1 \mu$G}  ($E_{\rm e}^{\rm br}\sim 50$ TeV), corresponding to the photon break energy $E_{\rm br}^{\rm IC}$ of $\sim 1$~TeV.

The observed X-ray synchrotron emission in young supernova remnants (e.g.~Vink et al.~\cite{Vink}) and the optical radiation in certain hotspots of radiogalaxies (e.g.~Brunetti et al.~\cite{BrunettiMack}) can support an interpretation based on the existence of unbroken electron spectra in the extended energy range. We presume that the conditions for efficient particle acceleration are achieved in peripheral radio relics because of the double shock, which allows us to expect such extensive electron spectra with a synchrotron turnover corresponding to $\gamma_{\rm e}^{\rm br}{m_{\rm e}c^2} \sim$~few~TeV. Since the synchrotron emission from monoenergetic electrons with $\gamma_{\rm e}$  peaks around $\nu_{\rm peak} \approx 0.3 (3 e B/ 4 \pi m_{\rm e}c)\gamma_{\rm e}^2$, then for electrons within the energy range $\sim\gamma_{\rm e}^{\rm br}\sim 10^8$ it corresponds to the break frequency $\nu_{\rm br,\,peak}\sim$~few $10^{15}$~Hz (for $B \sim 0.$1~$\mu$G), whose emission
is beyond the observational limit. Therefore we assume that for observed peripheral relics with straight spectra, this trend continues up to the frequency $\nu_{\rm br,peak}$. For the sake of illustration and to present a very approximate synchrotron flux estimation for Coma-like relics, we extrapolate the radio spectrum of a relic in Coma (cf. Fig.~1 in Atoyan \&~V\"olk~\cite{Atoyan};  Thierbach et al.~\cite{Thierbach}). The synchrotron emission generated by electrons with maximal energies ($\gamma_{\rm e}\sim10^8$) yields the value of flux, $\nu_{\rm syn} F_{\rm syn}$, of   $\sim10^8$~Jy\,Hz; for this very rough estimation we took the radio spectral index $\sim 1$. These electrons, which are responsible for the non-observable synchrotron emission in the EUV range, scatter the relic photons to the energy range of$\sim 1$~TeV. For the magnetic field $B \sim 0.1$~$\mu$G, the expected flux of IC emission $\nu_{\rm \gamma} F_{\gamma}$ roughly falls within the range ($10^{10} \div 10^{11}$)~Jy\,Hz, which is consistent with the expectation delineated in Fig.~\ref{Figure02}. 

In discussing the gamma-ray spectra below, we consider the energy range limited by $E\leq 10$~GeV. At the same time we emphasize that the expected shape of the concave spectrum is due to the physical reasons described above~as follows:
        \begin{itemize}
\item very efficient injection of relativistic protons, which are always expected to be around in this region;
\item their efficient reacceleration due to a double shock; and
\item $\rho_{\rm gas}$ strengthened by a 
forwarding shock; numerical simulations predict gas clumping (Vazza et al.~\cite{Vazza}; Vazza et al.~\cite{Vazza:a}; Eckert et al.~\cite{Eckert}).
        \end{itemize}

The possibility of gamma-ray spectral signature for $E \geq 100$~MeV, and the factors controlling this effect are the subject of a follow-up analysis. The parameters presumably playing a major role in shaping the spectrum are:   the value of the magnetic field, and   the relative acceleration efficiency of protons to electrons in the considered region.
      
Below we concentrate primarily on the dependence of the gamma-ray spectral structure on the magnetic field. 

First, we stress the energy ranges and timescales required in the radio relic model. For the Lorentz factors of electrons of  $3\times 10^{3} \div 10^{7}$, the cooling time via IC is $10^{9}$~y and $10^{5}$~y, and the average energies obtained by microwave photons are  $\approx 10^{3}\div 10^{10}$~eV, respectively. 
\medskip 

We limit ourselves to comparing gamma-ray emissions produced by two processes occurring in the location of radio relic at an edge of galaxy cluster and within the halo volume of another cluster owing to IC radiation of primary electrons and to   $\pi^{0}$ decay coming from pp interaction.

We expect that for a certain energy value $E$, 
the two fluxes will be comparable, and thus within the cooling timescales for electrons, the gamma radiation spectrum should reveal the structure  
 (a sort of intermittent phenomena),  
 indicating the presence of  both types of relativistic particle populations.

The gamma-ray flux produced by interaction of CR protons with ambient relic/halo nuclei is given by
        \begin{equation}
F=\frac{1}{4{\pi}D^2}\int{\!dV}\int\limits_{E_{\rm g}}^{\infty}\!\! dE\:q(r,E)
\label{eq:01}
        ,\end{equation}
where $V$ is the relic volume and $D$ is the luminosity distance.
To calculate the value of gamma-ray emissivity from $\pi^{0}$ decay, we used the formalism given in Pfrommer \& Ensslin~\cite{Pfrommer}, where
        \begin{eqnarray}
q_{\pi^0}(r,E)&=&\sigma_{\rm pp}c\,n_N(r)\;\xi^{2-\alpha_\gamma}\frac{n_p(r)}{\rm GeV}\frac{4}{3\alpha_\gamma}
\left(\frac{m_{\pi^0}c^2}{\rm GeV}\right)^{-\alpha_\gamma}\times\nonumber\\
&&~~\times\left[\left(\frac{2E}{m_{\pi^0}c^2}^{}\right)^{\delta\gamma}+\left(\frac{2E}{m_{\pi^0}c^2}\right)^{-\delta\gamma}\right]^{\alpha_\gamma/\delta\gamma}.
\label{eq:02}
        \end{eqnarray}
Here $\alpha_\gamma=\alpha_{\rm p}=\alpha_{\rm e}-1$ is $\gamma$-ray spectrum index, $\delta_{\gamma}=0.14\alpha^{-1.6}_\gamma + 0.44, \sigma_{\rm pp}=32\times(0.96 + e^{4.4-2.4\alpha_\gamma})\:$ mbarn, 
$m_{\pi^0}c^2/2\cong 67.5$~MeV, $\xi=2$, $n_N(r)=n_{\rm th}(r)/(1-0.5X_{\rm He})$ $(X_{\rm He}=0.24)$, where $n_{\rm th}(r)$ is the thermal electron density and $n_{\rm p}(r)$ is rescaled number density of CR protons (cf. Eq.~(8) Pfrommer \& Ensslin~\cite{Pfrommer}) expressed by $n_{\rm th}, \alpha_{\rm p}, X_{\rm p}$. 

\begin{figure}[htb]
\vspace{-1mm}
                \begin{center}
\includegraphics[totalheight=5.5 cm]{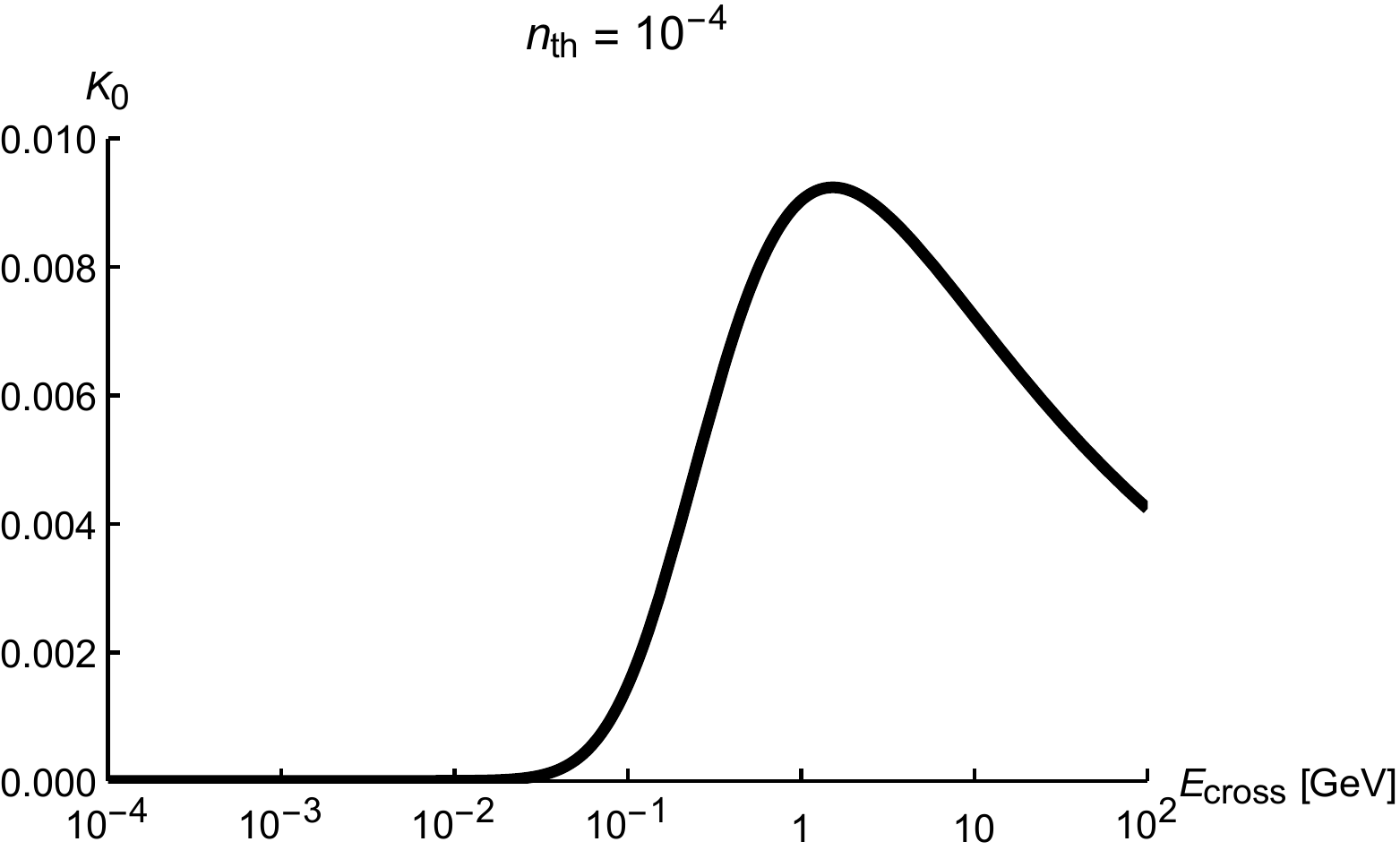}
                \end{center}
\vspace{-4mm}
\caption{Values of $K_{\rm 0}$ as a function of $E_ {\rm cross}$,
 i.e. energy at which
$q_{\pi}$ flux is equal to $q_{\rm IC}$ flux.}
\label{Figure03}
\vspace{-1mm}
        \end{figure}

The contribution to the gamma-ray flux coming from IC-scattered photons on primary electrons expressed by emissivity $q_{\rm IC}$ is given as
        \begin{eqnarray}
q_{\rm IC}(r,E)&=&\left[\frac{m_{\rm e}c^2}{\rm GeV}\right]^{1-p}\!n_{\rm e}(r)\:\frac{8\pi^2r_{\rm e}^2}{h^3c^2}\:F(p)(kT)^2
            \left[\frac{E}{kT}\right]^{\frac{-p+1}{2}},\nonumber\\
&&
\label{eq:03}
        \end{eqnarray}
where
        \begin{equation}
F(p)=2^{p+3}\frac{p^2+4p+11}{(p+3)^2(p+5)(p+1)}\Gamma\left[\frac{p+5}{2}\right]\xi\left[\frac{p+5}{2}\right]
\label{eq:04}
        \end{equation}
and
        \begin{equation}
p\equiv\alpha_{\rm e}.
        \end{equation}
The density $n_{\rm e} (r)$, of CR electrons  is often expressed by
        \begin{equation}
n_{\rm e}(r)=n_{\rm e}^0(r)\left(\frac{m_{\rm e}c^2}{\rm GeV}\right)^{-1+p}
\label{eq:06}
        ,\end{equation}
which is related to the electron energy distribution in two equivalent notations as follows:
        \begin{equation}
N(\gamma)d\gamma=n_{\rm e}^0\gamma^{-p}d\gamma~~\mbox{or}~
N_{\rm e}(E, r)=\frac{n_{\rm e}(r)}{{\rm GeV}}\left(\frac{E}{\rm GeV}\right)^{-p}.
\label{eq:07}
        \end{equation}
These electrons are scattered off the CMB photons in the blackbody temperature $T = 2.7$~K. 
 The analysis of  gamma-ray spectrum, which is the combination of both processes, reduces to following up the competition of these mechanisms, which in turn appears to depend 
on the ratio $K_{0}= n_{\rm e}/n_{\rm p}$. This quantity is in general not constrained by cosmic plasma physics and is often assumed as a constant. Nevertheless, its value is critical for the $q_{\pi}/q_{\rm IC}$ ratio and, in general, changes with energy. Because of the CRe energy losses, the number density ratio $K_{0}$ decreases with particle energy. In a zero-order approximation, if $K_{0}(E) = \mbox{const}$ were valid, one could ask at what energy the both emissivities are comparable. Thus, a comparison of both processes with the assumed fixed values of index~p and $n_{\rm th}$ in $q_{\pi}/q_{\rm IC} \propto n_{\rm th} K_{0} f(E)$ could lead to the possibility of estimating the  relation $K_{\rm 0}(E_ {\rm cross})$, see Fig.~\ref{Figure03}, and next allow us to evaluate the relative acceleration efficiency of electrons and protons. However, 
at energies around 1 GeV such an approach is not valid, as the electron losses through the IC process are significant here (energy distributions for e and p are not ``parallel") 
and thus $K_{0} = K_{\rm 0}(E) \ne \mbox{const}$ (e.g. Beck \& Krause~\cite{BeckKrause}). 
Therefore, below we concentrate on the input observables that are constrained by observational radio data and their implicit dependence on magnetic field $B$. To do this, we adopt  any necessary radio observables from the literature data for the examined cluster. Thus, in the 
further, $n_{\rm e}$, $n_{\rm p}$, $n_{\rm th}$ and the proton spectral indices are taken according to the analysed cluster case (referring to Reimer et al.~\cite{Reimer} or Siemieniec--Ozieblo \& Pasternak~\cite{Siemieniec--OziebloPasternak}, respectively). 
The procedure consists in attaching the values of variables resulting from the radio observation to the emissivities to reduce the number of parameters, i.e. the apparently high degree of degeneracy of the total gamma-ray flux.

\section{Two examples of a putative gamma spectrum and its dependence on the magnetic field}

The purpose of this analysis is to show that for both radio halo and radio relics in the galaxy clusters, a new (intermittent) feature is expected to appear in the gamma-ray spectrum in MeV energy range of Fermi--LAT observations. We also suppose that the energy at which this signature occurs is correlated with the value of the magnetic field 
 (corresponding to the place where the emission comes from)  via the relation of gamma-ray flux versus magnetic field. This relationship is by no means evident from the above formulae. The field $B$ affects the value of gamma-ray flux  via $n _ {\rm e} $ and/or $ n _ {\rm th} $. In the first case, $B$ can be regarded as a link in modelling the observed synchrotron radiation $F=F_0(n_{\rm e}(B))\nu^{-\alpha}$, permitting agreement between the adopted model of  radio emissivity and the observables that involve $B$. While $n _ {\rm th} $ can be associated with $B$ through the magnetic field simulation to consistently match its spatial profile with RM observations, e.g. Bonafede et al.~\cite{Bonafede} (discussed below). 
Observational confirmation of this effect would be of particular importance for estimating the value of intergalactic magnetic field. This new method would be especially useful in the case of radio relics, which occur mostly at the edges of clusters. To illustrate this effect, we consider the expected gamma-ray spectra for the two cases well known in the literature, i.e. the radio halo of Coma cluster and the relic of A~2256 cluster.

For the Coma cluster, the basic observables concerning relativistic electrons ($n_{\rm e}, \alpha_{\rm e}$) come from radio observations (Thierbach et al.~\cite{Thierbach}) while another observational quantity, X-ray emissivity, allows us to roughly estimate the density of thermal particles, $n_{\rm th} \sim 10^{-3}$~cm${}^{-3}$. This value corresponds to the thermal energy density $u_{\rm th} \cong 3.8\times 10^ {- 11}$~erg\,cm$^{- 3}$. Following Pinzke \&~Pfrommer~\cite{Pinzke}, we take the electron spectral index $\alpha_{\rm e} = 3.6$, which correctly reproduces the radio flux density  $I_ {\rm syn} \propto\nu^{(1-p)/2}\exp\left(-\sqrt{\nu/\nu_{\rm s}}\right)$ and $\nu_{\rm s} = 0.44$~GHz from Thierbach et al.~\cite{Thierbach}. The hypothetical magnetic field value  $B \sim 0.1 \div 1$~$\mu$G fits well within the range that allows us to reproduce the observed radio flux of synchrotron radiation. The above quantities can be used to calculate the expected value of $q_{\rm IC}$, according to the value of $B$. For the $q_{\pi^0}\propto n_{\rm p} \times n_{\rm th}$,  the  CRe index is known, and thus (assuming a common acceleration mechanism) the spectral index of protons is also known. Then  to eliminate $n_{\rm p}$, instead of the parameter $K_{0}$, we use the parameter $X = {n_{\rm p}}/{n_{\rm th}}$ (i.e. a rescaled version of $X_{\rm p} =  {\epsilon_{\rm p}}/{u_{\rm th}}$). Given that the proton injection to acceleration area depends highly on gas density, we assume that $X \sim \mbox{const}$. Furthermore, in the cluster outskirts we can perceive its high value as a result of a relatively high-energy content in CRp relative to  $n_{\rm th}$ (see e.g. Brunetti \&~Jones~\cite{BrunettiJones}). 

\begin{figure}[htb]
                \begin{center}
\includegraphics[totalheight=5.5 cm]{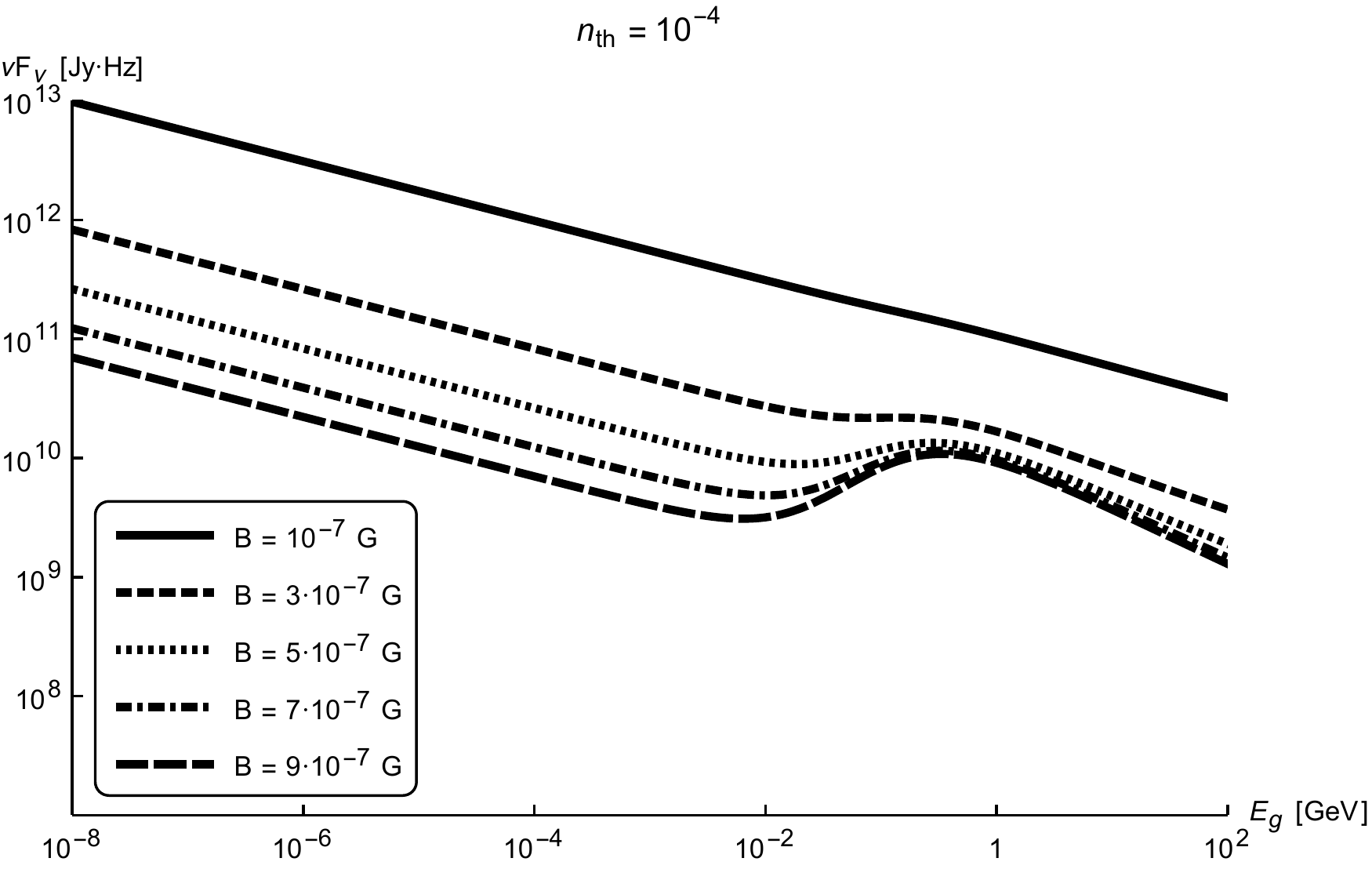}
                \end{center}
\vspace{-4mm}
\caption{Expected gamma-ray spectrum for thermal plasma density, $n_{\rm th} = 10^{-4}$~cm${}^{-3}$ of the Coma cluster. The sum of  IC and $\pi^0$ fluxes is calculated for the field strength values given in the box and presented from upper to lower curves. The physical observables  $n_{\rm e}, \alpha_{\rm e}$ (used to fit the radio data) and $X_{\rm p}=0.28$ are taken from Reimer et al.~\cite{Reimer}.}
\label{Figure04}
        \end{figure}

\begin{figure}[hbt]
        \begin{center}
\includegraphics[totalheight=5.5cm]{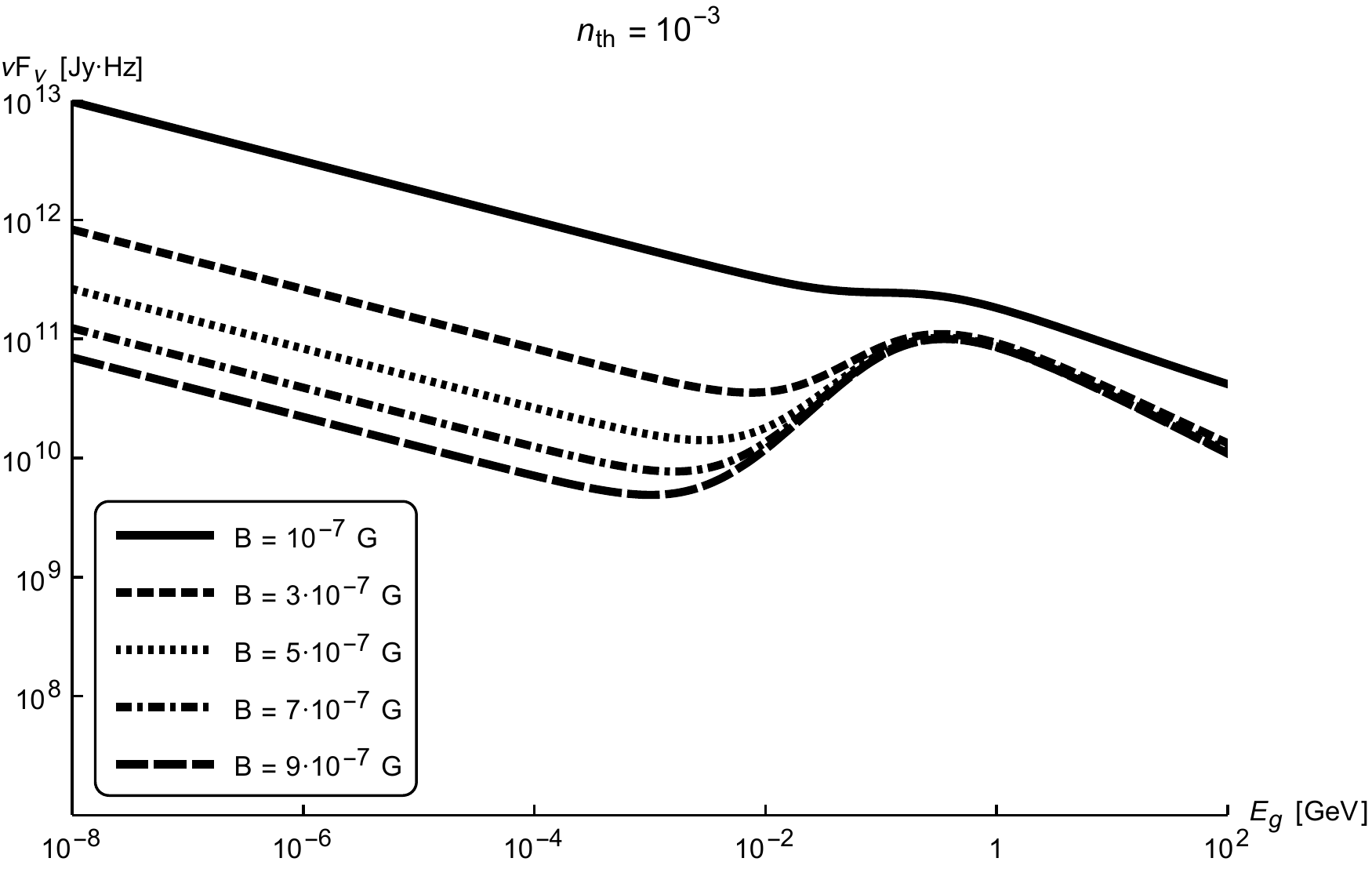}
                \end{center}
\vspace{-4mm}
\caption{Same as Fig.~\ref{Figure04} but fluxes are calculated for thermal density $n_{\rm th} = 10^{-3}$~cm$^{-3}$.}
\label{Figure05}
        \end{figure}

\begin{figure}[hbt]
        \begin{center}
\includegraphics[totalheight=5.5cm]{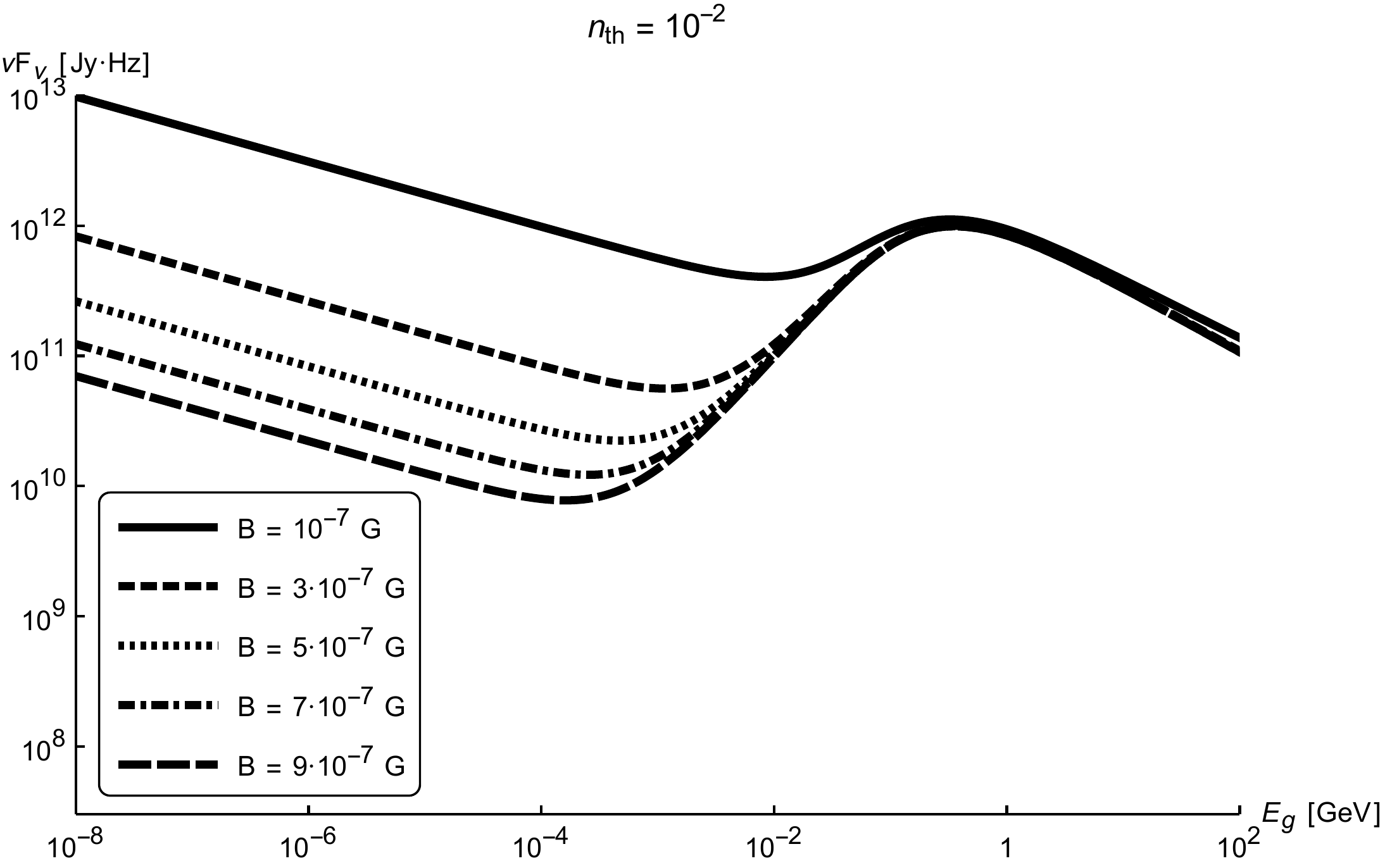}
                \end{center}
\vspace{-4mm}
\caption{Same as Fig.~\ref{Figure04} but fluxes are calculated for thermal density $n_{\rm th} = 10^{-2}$~cm$^{-3}$.}
\label{Figure06}
        \end{figure}

Thus taking the hadronic spectral index $\alpha_ {\rm p} = 2.5$ and the postulated ratio of number density  of protons to the thermal gas $X_{\rm p}=0.28 $ (for the sake of comparison with the results of Reimer et al.~\cite{Reimer}) allows us first to calculate the needed density of  $n_{\rm p}$ and $n_{\rm e}$ and then to compute the emissivity of both processes supposed to produce gamma radiation. According to Eqs.~(\ref{eq:01}),~(\ref{eq:02}), and~(\ref{eq:03}), the total gamma-ray flux for different values ​​of the magnetic field is shown in Fig.~\ref{Figure04}, which, together with Figures~\ref{Figure05} and~\ref{Figure06}, visualize how the spectral structure changes with decreasing thermal density.
Despite the dominant role of the IC process in the energy range $10^{- 4} \div 10^{-2}$~GeV, this spectral signature still is visible for the $B$ fields $<1$~$\mu$G. The value of energy, $E_ {\rm cross}$ at which both the gamma fluxes, i.e. for IC and $\pi^0$ processes are comparable, changes with magnetic field $B$ and density of thermal medium $n_ {\rm th}$ variations in the ICM. 

\begin{figure}[htb]
        \begin{center}
\includegraphics[totalheight=5.5cm]{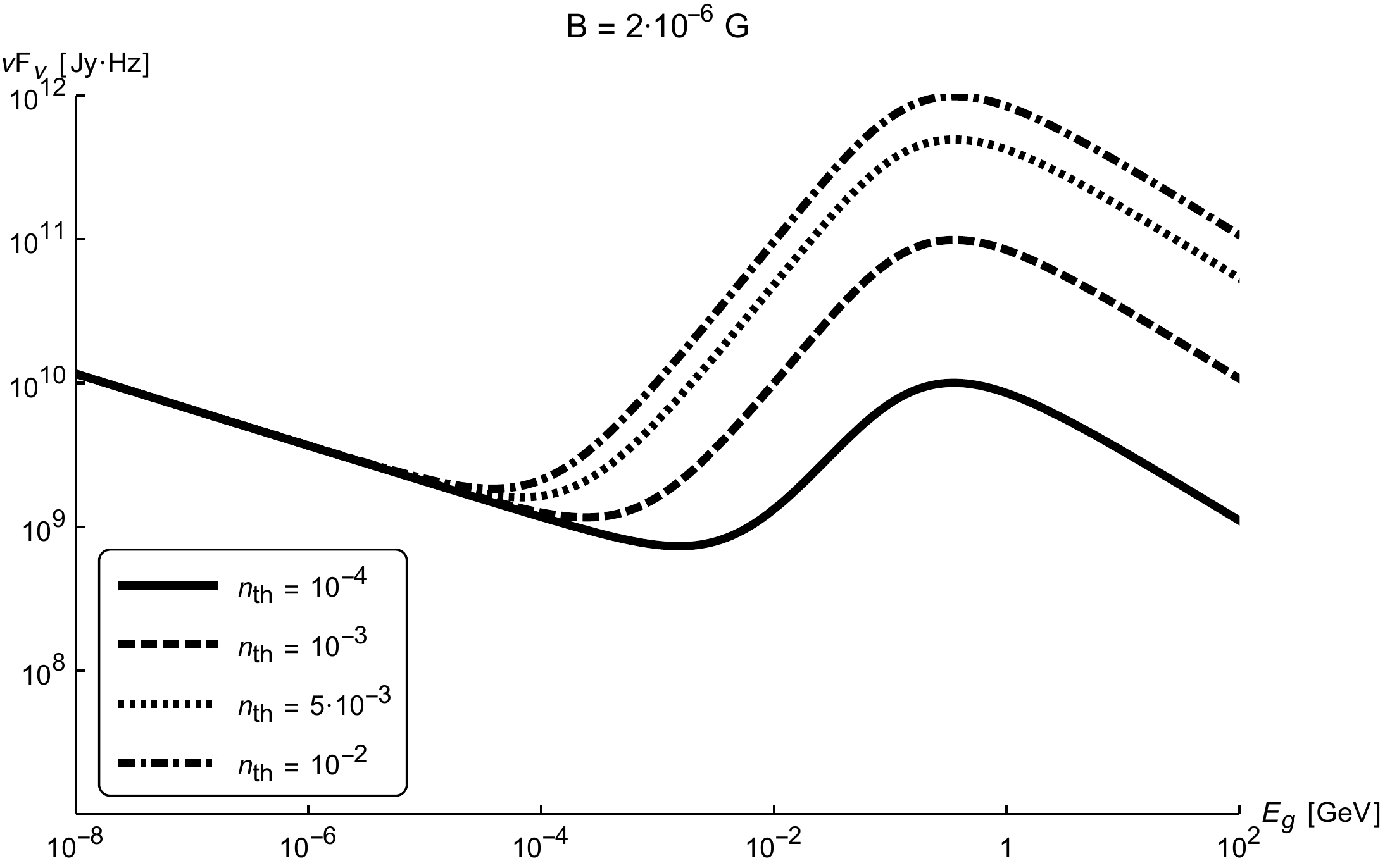}
                \end{center}
\vspace{-4mm}
\caption{Total  gamma-ray fluxes for different  thermal plasma densities shown for the field strength $B = 2~\mu$G.}
\label{Figure08}
        \end{figure}

\begin{figure}[htb]
        \begin{center}
\includegraphics[totalheight=5.5cm]{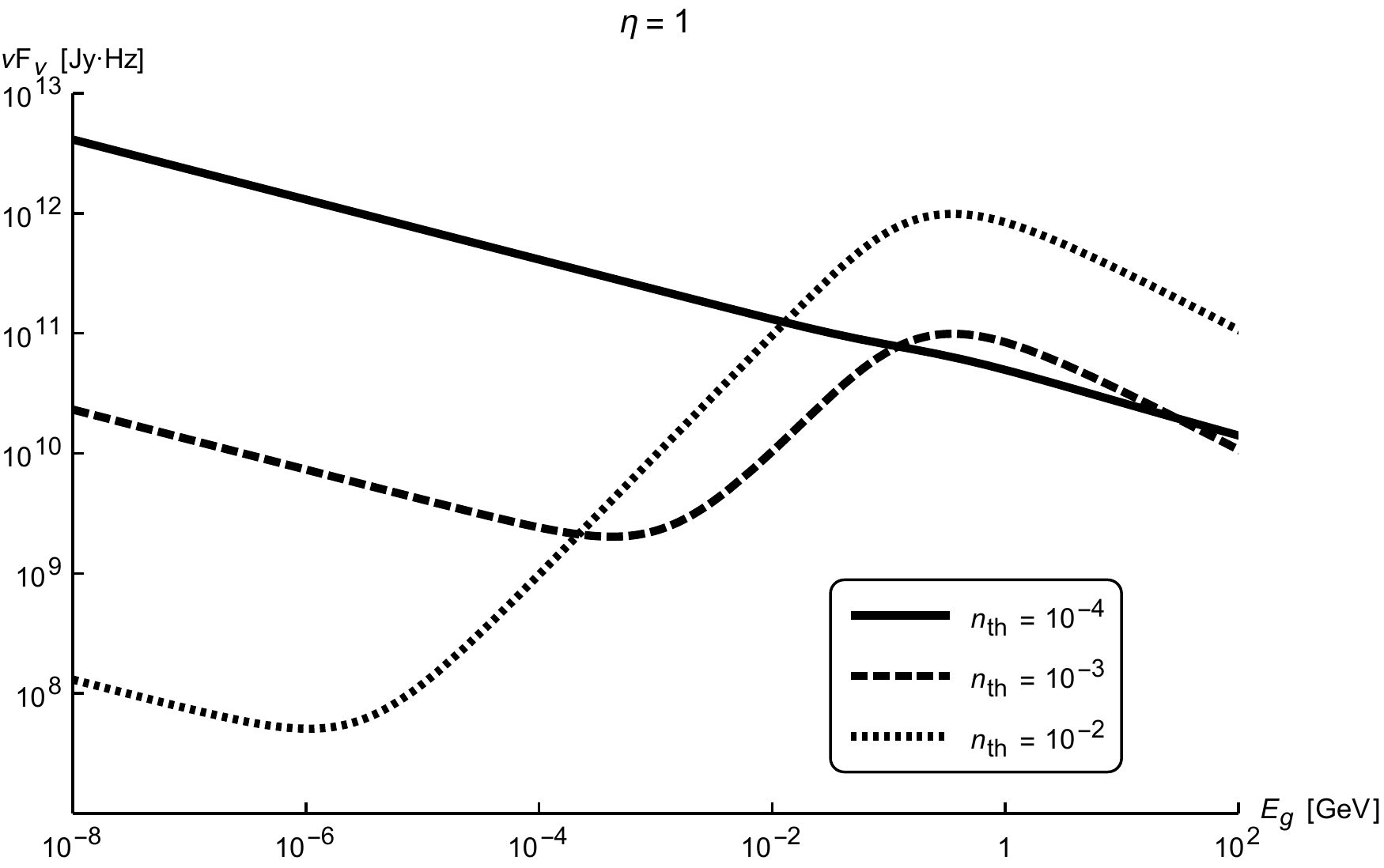}
                \end{center}
\vspace{-4mm}
\caption{Profiles of  gamma-ray fluxes shown for different  thermal plasma densities  with the magnetic field strength eliminated via Eq.~\ref{eq:08}, for 
$\eta = 1$.
}
\label{Figure09}
        \end{figure}

The $E_ {\rm cross}$ value decreases with increasing value of $n_ {\rm th}$ and with the value of $B$ as well, while the flux $F_ \nu(E_ {\rm cross})$ value increases with $n_ {\rm th}$,  (Fig.~\ref{Figure08}). It should be noted that these two variables, i.e. $B$ and $n_ {\rm th}$,  both depending on $r$, are not be treated as independent quantities. They usually scale according to the phenomenological formula
        \begin{equation}
\langle B(r)\rangle=B_0
\left(
\frac{n_{\rm th}(r)}{n_0}
\right)^\eta
\label{eq:08}
        ,\end{equation}
where
$\eta\in (0.1\div 1)$ and $n_0=3.4\times 10^{-3}$~cm${}^{-3}$.
Theoretically, the above relationship originates from the conservation law for magnetic flux during the gravitational collapse of a magnetized structure, when $n_ {\rm th} \sim B^a$. However for the frozen-in magnetic field compression required by the power-law relation with plasma density, the average field~$B$ strongly depends on multiple parameterizations. The most important variables are related to compression symmetry and magnetic field amplification processes of different kinds during the cluster evolution. They play a significant role at the cluster outskirts because of constant accretion and mergers. Thus accepting the phenomenological approach by applying the above scaling relation, while it imposes  some uncertainty, seems to be the only reliable way to interrelate these variables.  
 The choice of right values of ​$​\eta$ and $B_ 0$ is usually carried out by correlating the different observables (e.g.~RM versus X-ray brightness); see for example~Donnert et al.~\cite{Donnert:a} and Bonafede et al.~\cite{BonafedeBruggen}. To illustrate how to remove the degeneracy between $n_ {\rm th}$ and $B$, we take, following Donnert et al.~\cite{Donnert:a}, $\eta \sim 1$ (Fig.~\ref{Figure09}). 
The value of energy at which the considered structure appears now depends only on the value of $n_ {\rm th}$, which is a function of radial position of the examined  region. This position can therefore be assumed to be the very place of structure occurrence that is associated with both $n_ {\rm th}$ and $B$. This seems to be very advantageous for clusters because the detector spread function at $\sim$~ 1 GeV range is typically of the order of the cluster size. On the other hand, the knowledge of $E_ {\rm cross}$ value gives us a chance to estimate the value of magnetic field $B$, provided the scaling law for a given cluster is properly established. In general, by adopting such a scaling procedure between the cluster thermal density and its magnetic field, we can infer the following relation from the analysed requirement: 
\begin{equation}
Q:=\frac{q_{ \pi}}{q_{IC}}(E_{\rm cross})\simeq1,
\end{equation}
\begin{equation}
Q(E_{\rm cross})= \frac{n_{\rm th}n_{\rm p}}{n_{\rm e}^0 f(p)}= \frac{X n_{\rm th}^2}{f(p) n_{\rm e}^0(B)},
\end{equation}
and finally
\begin{equation}
Q(E_{\rm cross})= \frac{X B^{g(\eta)}}{\tilde{f}(p) n_{\rm e}^0(B)}
,\end{equation} 
where $g(\eta)$, $\tilde{f}(p)$ and $f(p)$ are 
 the following functions of  $\eta$ and index $p$, respectively:

\[
g(\eta)= \frac{2}{\eta},~~\tilde{f}(p)=f(p)\left(
\frac{B_0^\frac{1}{\eta}}{n_0}
\right)^2
\] 
and
        \begin{eqnarray}
f(p)&=&\frac{A}{C}
\frac{(p-1)F(p){\cal B}(\frac{p-3}{2},\frac{4-p}{2})\left[\frac{E_{\rm cross}}{kT}\right]^\frac{1-p}{2}}
{(p-2)\xi^{3-p}\left[\frac{m_{\pi^0}c^2}{{\rm GeV}}\right]^{-\alpha_\gamma}}\times\nonumber \\ 
&&\times  
\left[
\left(\frac{2E}{m_{\pi^0}c^2}^{}\right)^{\delta\gamma}+\left(\frac{2E}{m_{\pi^0}c^2}\right)^{-\delta\gamma}
\right]^{-{\alpha_\gamma}/{\delta_\gamma}},\nonumber 
        \end{eqnarray}
where ${\cal B}$ it is the beta function, 
\[
A= \frac{8\pi^2 r_{\rm e}^2}{h^3 c^2}(kT)^2
\]
and
\[
C=4\frac{\sigma c}{m_{\rm p}c^2}
\frac{(2-1.25 X_{\rm He})(k T_{\rm gas}/{\rm GeV})}
{(1-0.5 X_{\rm He})^2},
\]
$\alpha_\gamma=p-1$, $\delta_\gamma=0.14\alpha_\gamma^{-1.6}+0.44$ and $F(p)$ is given by eq.~\ref{eq:04}. 
On the other hand, from the observed radio spectrum we have the value of  $F_{\rm 0}$, related to $B$ via $ n_{\rm e}$ (eq.~\ref{eq:06}), which finally gives
\[
n_{\rm e}^0=\frac{4\pi D^2 F_0}{\kappa V}B^{-\frac{p+1}{2}},
\]
where
\[
\kappa=\frac{\sqrt{3}e^3}{m_{\rm e}c^2(p+1)}\Gamma
\left(\frac{p}{4}+\frac{19}{12}\right)
\Gamma
\left(\frac{p}{4}-\frac{1}{12}
\right)
\left[
\frac{2\pi m_{\rm e}c}{3e}
\right]^\frac{1-p}{2}.
\]
Knowledge of gamma-ray flux for $E_{\rm cross}$  plus the above formulae allow us to  constrain the required variables (for a~given $X_{\rm p}$), i.e. $B$ and $\eta$ (or for previously fixed $\eta$: $B$ and $X_{\rm p}$). Similarly, as in the case of hadron models for radio halos where the combination of radio and gamma-ray data  provides a constraint on $B$, see e.g. Brunetti \& Jones~\cite{BrunettiJones}, we have here the combination of radio flux and spectral shape of gamma-ray flux. Thus, applying the above procedure, via scaling $n_{\rm th}$ versus $B$, one obtains an independent limitation on $B$.

\begin{figure}[htb]
        \begin{center}
\includegraphics[totalheight=5.5cm]{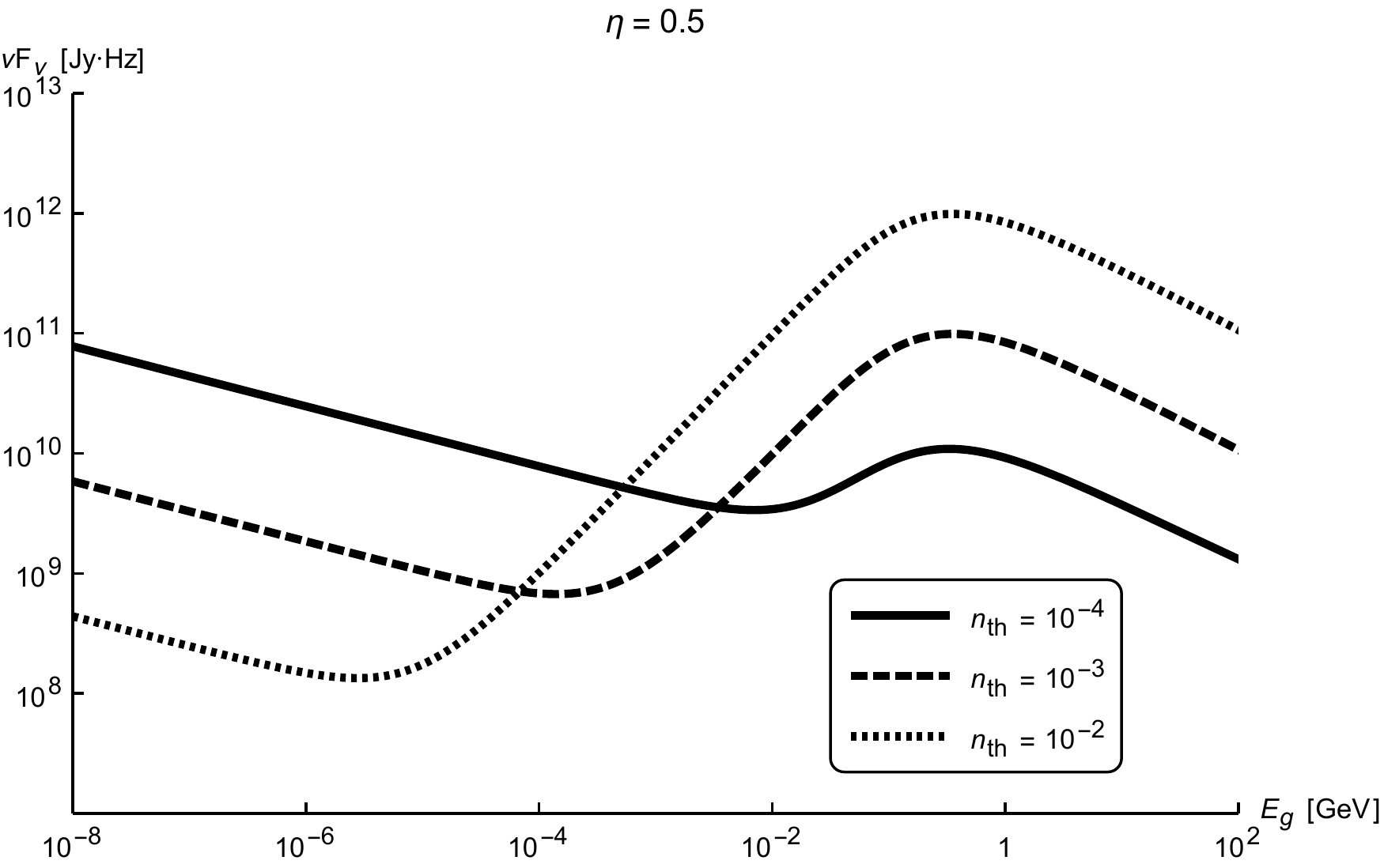}
                \end{center}
\vspace{-4mm}
\caption{Same as Fig.~\ref{Figure09}, fluxes calculated for $\eta = 0.5$.}
        \label{Figure10}
        \end{figure}

\begin{figure}[htb]
        \begin{center}
\includegraphics[totalheight=5.5cm]{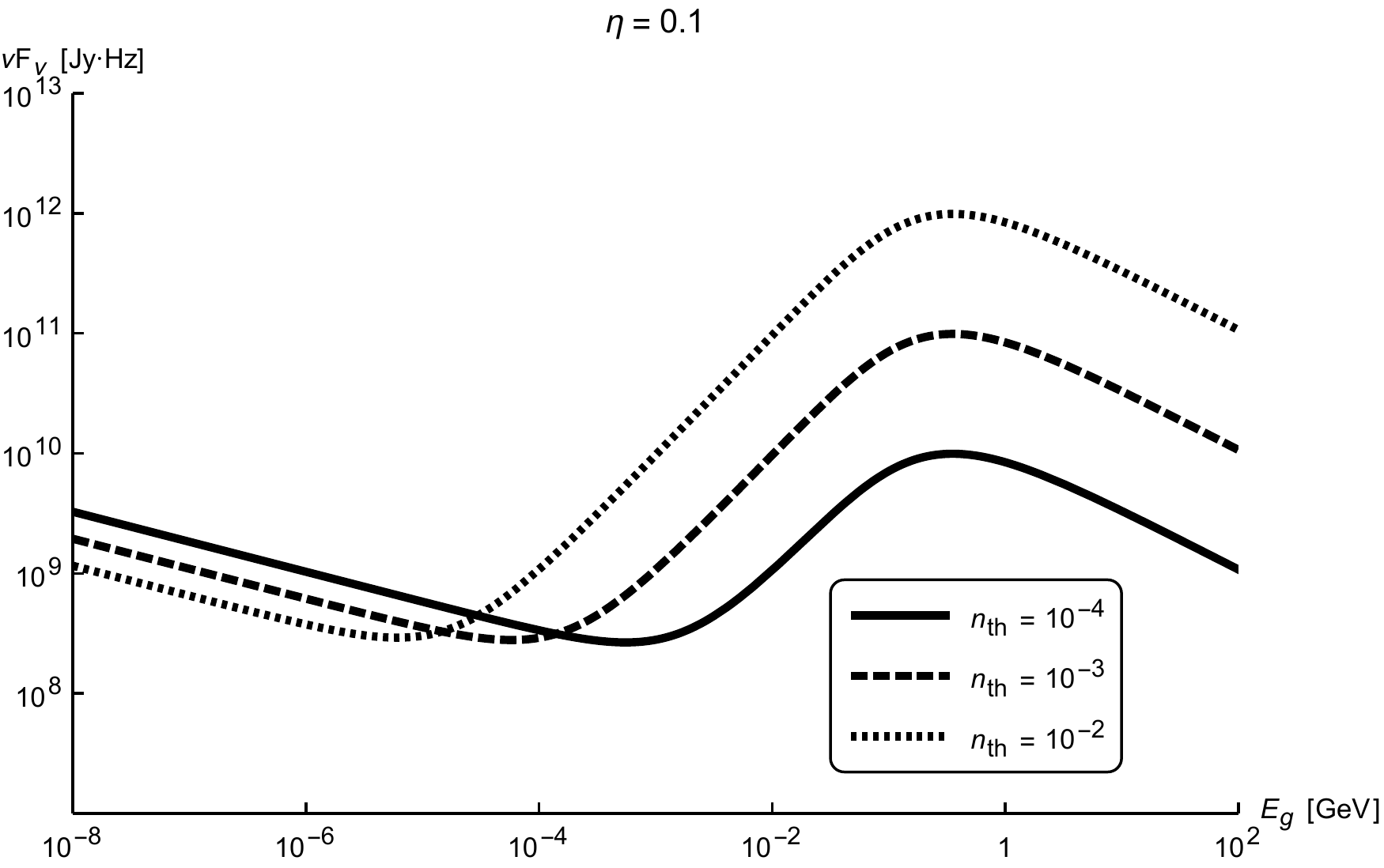}
                \end{center}
\vspace{-4mm}
\caption{Same as Fig~\ref{Figure09}, fluxes calculated for $\eta = 0.1$.}
\label{Figure11}
        \end{figure}

Also Figures~\ref{Figure10} and \ref{Figure11} show a moving point $E_ {\rm cross}$ in the gamma-ray spectrum for other values ​​of the $\eta$ parameter in the scaling law $n_ {\rm th} (r)$ versus $B(r)$.
            

Another kind of diffusive radio emission especially convenient to compare with the Coma cluster halo is the relic of the A~2256 cluster. Here we have a radio relic whose emission comes from the volume V, which is comparable to the Coma halo volume. The radio fluxes for both diffusive objects are also comparable, as for the halo in Coma $F_\nu \sim 2$~Jy, at a frequency of $\sim 400$~MHz, while for the relic in A~2256, $F_\nu \sim 1.4$~Jy, at a frequency of 350~MHz.
In the case of the above relic, the basic observables result from a method that is very different than previous methods; this  is briefly described below. We expect that the relic, which occurs at the edge of cluster,  provides particularly good conditions for observing the signature in the gamma-ray spectrum discussed above. Essentially, the value of the magnetic field decreases on the edge of the cluster, and because of this, the gamma-ray component due to the IC process acting on a primary population of freshly accelerated electrons becomes relatively bright.
The observables necessary to calculate a relevant gamma-ray flux were obtained in the papers quoted below. The approximate value of $n_ {\rm th} \sim 0.5\times 10^{- 3}$~cm$^{- 3}$ at the edge of the cluster is taken from Brentjens~\cite{Brentjens}. The density of proton and electron numbers  $n_ {\rm p} \sim 10^{-11}$~cm$^{-3}$,  $n_ {\rm e}\sim 3.6 \times 10^{- 12}$~cm$^ {- 3}$ and the value of the magnetic field $B \sim 0.05$~$\mu$G were calculated in the paper Siemieniec--Ozieblo \& Pasternak~\cite{Siemieniec--OziebloPasternak}, basing on the knowledge of 
        \litem{$-$}  flux values of synchrotron radiation and non-thermal X-ray emission (IC) and the spectral index characterizing the relic radio emission $\alpha = 0.8$,
and        \litem{$-$} the postulate of comparable energy density in hadron and lepton fractions of CR, expressed through the value of the parameter $K = \rho_{\rm p} / \rho_{\rm e}\sim 1$.

\begin{figure}[htb]
        \begin{center}
\includegraphics[totalheight=5.2cm]{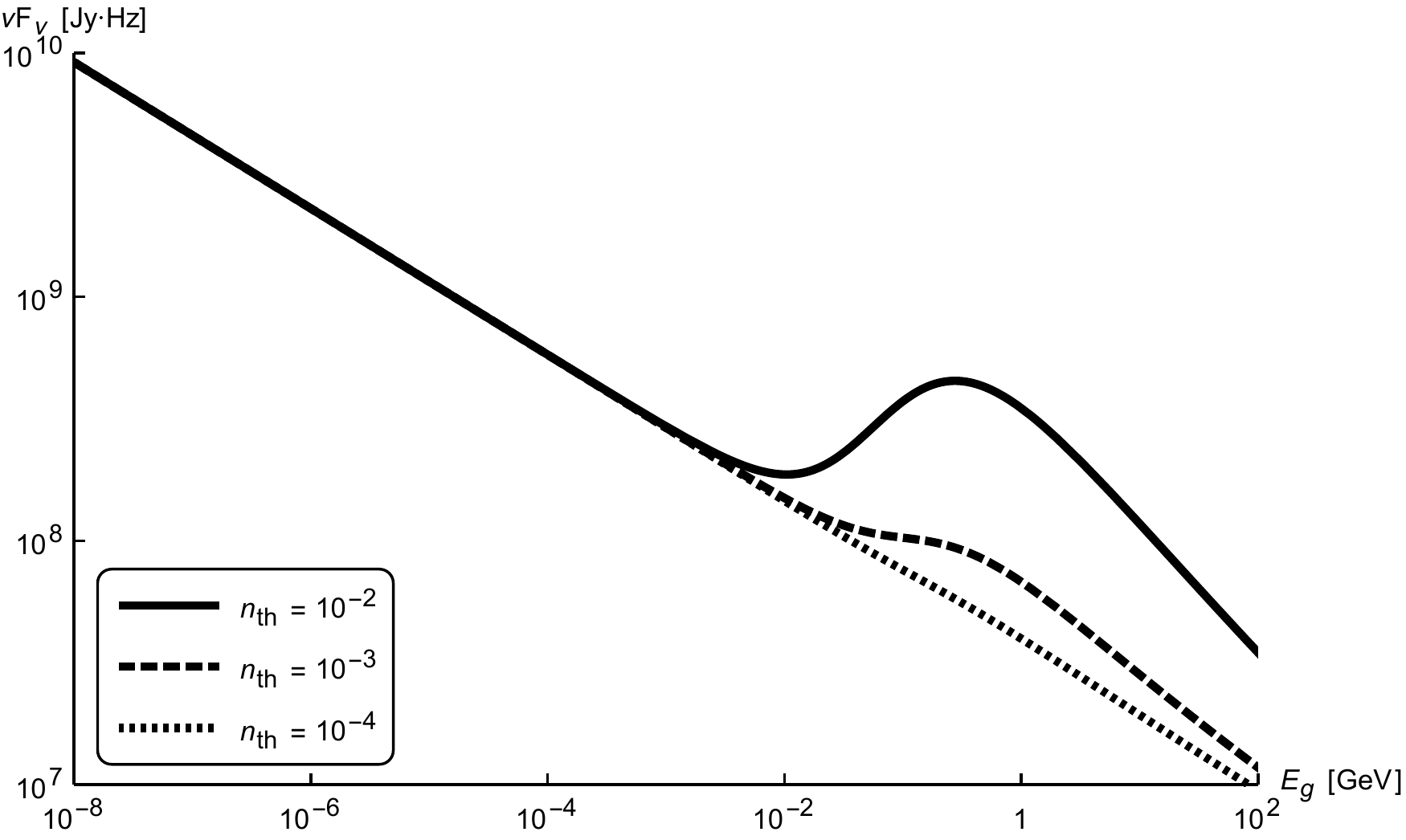}
                \end{center}
\vspace{-4mm}
\caption{Combined  gamma-ray emissions for a radio relic at the outskirts of A~2256 galaxy cluster  for different values of  $n_{\rm th}$, where the values of $n_{\rm e}$, $n_{\rm p}$  were inferred from radio and HXR observations (Siemieniec--Ozieblo \& Pasternak~\cite{Siemieniec--OziebloPasternak}).}
\label{Figure12}
        \end{figure}

It should be noted that the values of $B$ listed above and $n_{\rm e}^0$ factor $\sim 10^{- 4}$~cm$^{- 3}$ (in the formula (\ref{eq:07}) on $N(\gamma)$; $p =  1 + 2\alpha$) are comparable with the values of the observables obtained from the   method previously mentioned, used in Brentjens~\cite{Brentjens}. Taking the above into account  these values, which result from the two different approaches, seem to be very reliable. In analysing the structure of gamma-ray spectrum for the relic in A~2256, the only ``free" parameter, as opposite to the Coma case, is the value $n_ {\rm th} (r)$ in the relic. The values for $B$ and $n_{\rm e}^0$ result from solving flux equations (1), (2), in Siemieniec--Ozieblo \& Pasternak~\cite{Siemieniec--OziebloPasternak}. Figure~\ref{Figure12} shows the same trend in the behaviour  of $E_ {\rm cross}$ as in the halo of Coma, implying that the structure within the gamma-ray spectrum depends on $n_ {\rm th}$. As before, with decreasing value of $n_ {\rm th}$ (for increasing $r$) we have an increasing value of $E_{\rm cross}$.

\section{Summary}

The growing interest in the physics of cluster edges, underlying recent attempts to have a better observational access to them is primarily due to much inadequate knowledge of that virial region that directly links the ICM and the outer IGM. The cluster outskirts are particularly important not only because the cluster mass distribution, essential for investigating structure formation evolution, is still unknown to date, but also from the accretion of matter along the filaments, which, on the one hand, leads to effects such as gas clumping and, on the other hand, to the shock occurring there. In fact, the gravitational energy transfer proceeds through the accretion of matter and the associated shocks. The energy is dissipated here to gas, plasma turbulences, magnetic field $B$, and CR acceleration. The main consequence of shock actions is  acceleration/reacceleration of particles, leading to the appearance of non-thermal IGM plasma component. This non-thermal component is crucial not only for the understanding of different radiative mechanisms but also in contributing to non-thermal pressure, which additionally could influence the out-of-equilibrium ICM state that is relevant to this region.
The flux of non-thermal radiation significantly depends on the strength of magnetic field frozen in the peripheral plasma. The present knowledge of the  $B$ field in the IGM is scanty and its estimated value often depends on the particular method used.
In general, many aspects of the $B$ field in clusters are still far from being understood, in particular, its role in the ICM and IGM environments connecting clusters with the large-scale web structure.
In this context, any new independent method of estimating the intensity and getting additional information on the IGM magnetic field would be highly desirable. Potentially, this could be achieved by a future detection of gamma-ray spectrum  from the edge of the cluster in which a radio relic is observed as well. At present, in some cases, we can  also use edge-on radio relics to estimate $B$ (via spectral aging), while this method cannot be applied to face-on relics.
So far, only the upper limit on gamma-ray flux was determined by FERMI. Lack of observations constrains the contribution of CRs and thus the CR pressure averaged over the entire cluster. However, there are a number of arguments  (e.g.~Hong et al.~\cite{Hong} and  Vazza et al.~\cite{Vazza}) to suggest that the CR pressure can rise at the cluster edges.
Referring  to the work of Hong et al.~\cite{Hong}, where protons are accelerated mainly in the outskirts (through infall shocks), in this paper 
we propose a scenario in which the CR acceleration is due to a double shock, i.e. a  shock involving both merger and accretion shocks.
Propagating in the direction of accretion shock, the outwards merger shock creates on the edge of the cluster favourable conditions for the appearance of an efficiently  accelerated population of electrons. Thus, exactly in this area, we expect an inter-shock gamma-ray emission arising both from the $\pi^{0}$ decay and the IC processes. However, since the value of $B$ field is small there, the interaction of electrons with CMB photons is more efficient than their interaction with the field $B$. Therefore, the contribution to gamma-ray emission from the IC process becomes relatively large and thus comparable with that of $\pi^{0}$ decay. 
Such a combination of both mechanisms should therefore create a gamma-ray  spectrum (on the timescale of the radio relic) revealing a characteristic time-dependent structure.
 As has been shown in the example of two often studied clusters whose observables were derived in previous papers, this feature is located in the photon energy range less than 1 GeV for small values of magnetic field, less than 1 $\mu$G. A large-scale simulation of gravitational structure suggests just  such a magnetic field in the transition zone from the ICM environment to the IGM filament milieu. 

The energy at which the inner structure of the spectrum becomes distinct scales with the value of peripheral magnetic field. Therefore, the knowledge of, for example the thermal gas profile of the cluster and the spectral position of a given gamma-ray feature, could be essential for an independent estimation of magnetic field values at the interface between the ICM and IGM.

Moreover, the potential capability is very promising to see the close neighbourhood of the $\pi^{0}$ bump and to find two similar spectral signatures on both sides of this bump in two different energy ranges: at $\lesssim $ 1 GeV via Fermi--LAT and $> 10$~GeV via future CTA instruments (see~Fig.~\ref{Figure02}). The Fermi and CTA overlapping in the extended energy range plus enhanced sensitivity and angular resolution should provide a~supplementary tool to constrain the acceleration scenario. These parallel observations will in the future enable us to overcome  degeneracy of gamma-ray spectrum, depending both on the magnetic field and relative acceleration efficiency for electrons and protons.

\begin{acknowledgements}

{We thank the referee for helpful and insightful comments. }

\end{acknowledgements}

\end{document}